\documentclass[journal]{IEEEtran}
\usepackage{cite}

\ifCLASSINFOpdf
\else
\fi
\usepackage{amsmath}
\usepackage{array}
\usepackage{fixltx2e}
\usepackage[justification=centering]{caption}
\usepackage[pdftex]{graphicx}
\usepackage{booktabs}
\usepackage{subfloat}
\usepackage{multirow}
\usepackage{array}
\usepackage{setspace}
\usepackage{bm}
\usepackage[caption=false,font=footnotesize]{subfig}

\usepackage{url}

\begin{document}
%
\title{Spec-ResNet: A General Audio Steganalysis Scheme based on a Deep Residual Network for Spectrograms}
%
%
%

\author{Yanzhen~Ren,~\IEEEmembership{Member,~IEEE,}
~Dengkai~Liu,~Qiaochu~Xiong,~Jianming~Fu,~and~Lina~Wang
\thanks{This work is supported by the Natural Science Foundation of China (NSFC) under  grant nos. 61872275, U1536114,  and U1536204 and the China Scholarship Council.}
\thanks{Yanzhen Ren, Dengkai Liu, Qiaochu Xiong, Jianming Fu and Lina Wang are with Key Laboratory of Aerospace Information Security and Trusted Computing, Ministry of Education, School of Cyber Science and Engineering, Wuhan University, Wuhan 430072, China. (e-mail: renyz@whu.edu.cn; dengkailiu@whu.edu.cn; emmaxiong@whu.edu.cn; jmfu@whu.edu.cn; lnawang@163.com; )} }

\maketitle

\begin{abstract}
The widespread application of audio and video communication technologies has resulted in compressed audio data flowing across the Internet and has made it an important carrier for covert communication. Many steganographic schemes have emerged for mainstream audio data compression, including AAC and MP3, with many steganalysis schemes being subsequently developed. However, these steganalysis schemes are only effective in the given embedding domain. In this paper, a general steganalysis scheme, Spec-ResNet (Deep Residual Network of Spectrogram), is proposed to detect steganography schemes in different embedding domains for AAC and MP3. The basic idea is that the steganographic modification of different embedding domains all introduce  changes in the decoded audio signal. In this paper, a spectrogram, which is a visual representation of the spectrum of the frequencies of an audio signal, is adopted as the input of the feature network to extract the universal features introduced by steganography schemes. Our deep-neural-network-based method, Spec-ResNet, can effectively represent steganalysis features, and the features extracted from different spectrogram windows are combined to fully capture the steganalysis features. The experimental results show that the proposed scheme achieves good detection accuracy and generality. The proposed scheme achieves higher detection accuracy for three different AAC steganographic schemes and MP3Stego compared to state-of-the-art steganalysis schemes based on traditional hand-crafted or CNN-based features. To the best of our knowledge, this audio steganalysis scheme based on spectrograms and a deep residual network  is the first to be published. The method proposed in this paper can be extended to the audio steganalysis of other codecs or audio forensics.
\end{abstract}


%
\IEEEpeerreviewmaketitle

\section{Introduction}
%
%
%
%

\IEEEPARstart{S}{teganography}  is the technique of concealing secret information in digital carriers, such as text, images, audio, and video, to facilitate covert communication. In many cases, these steganography schemes are used by malicious software or malicious organizations. Therefore, steganalysis technology, which is the countermeasure technique against steganography for detecting whether there is secret information in data that appear normal, has been extensively studied.

With the widespread application of various audio and video communication technologies, compressed audio is stored and spread across the Internet. Today, the most widely used audio compression standards are AAC (Advanced Audio Coding) and MP3 (MPEG-1 Audio Layer III). AAC plays a dominant role in various Internet communication applications and can achieve better sound quality than MP3 at similar bit rates; therefore, AAC is gradually replacing MP3. AAC is specified  as part of both the MPEG-2 and MPEG-4 standards\cite{iso2004iec}; therefore, it is used almost everywhere for mainstream audio applications. Today, AAC is the default audio coding format for many mainstream HDTV standards, hardware, mobile phones, and Internet communication applications\cite{aac} such as DVB (Digital Video Broadcasting), iTunes, the iPod, the  iPhone, the Sony PlayStation, YouTube, Twitter, Facebook, Youku, Tudou, Baidu Music, QQ Music, and Himalayan.

The ubiquitous storage and communication of audio compression data provides a huge covert communication carrier channel. Many steganography schemes based on AAC and MP3 have been proposed\cite{wang2011steganography,xu2009performance,wei2010controlling,wang2010steganography,zhu2010sign,zhu2011huffman,Luo2017Adaptive,
bazyar2014recent,yang2017adaptive,zhang2017steganography,petitcolas1998mp3stego,platt2004undermp3cover,achmad2008mp3stegz}.
There are three main embedding domains concerning the compression parameters of AAC: Modified Discrete Cosine Transform coefficients (MDCT)\cite{wang2011steganography}, the scale factor \cite{xu2009performance,wei2010controlling}, and Huffman coding \cite{wang2010steganography,zhu2010sign,zhu2011huffman}.
The basic principle behind these steganography schemes is in modifying the compression parameters to hide the secret information. Wang\cite{wang2011steganography} proposed to embed secret information by modifying the small value of quantized MDCT coefficients to achieve high imperceptibility. In \cite{wang2010steganography}, the least significant bit (LSB) of the escape sequences in AAC Huffman coding is modified with matrix encoding\cite{Fridrich2006Matrix} to embed secret information. In \cite{zhu2010sign}, the sign bit of the corresponding MDCT coefficient is modified to implement steganography, therein utilizing the fact that the sign bits of the MDCT coefficients in the signed codebook will be coded separately. In \cite{zhu2011huffman}, Zhu proposed to embed secret information by modifying the sections of the Huffman coding of AAC.
In \cite{Luo2017Adaptive}, Luo proposed an adaptive audio steganography scheme in the time domain based on AAC and Syndrome-Trellis coding (STC).
Except for \cite{Luo2017Adaptive}, the above steganography schemes  all modify the compression parameters of AAC. Although the embedding domains are different in each scheme,   the MDCT coefficients of each AAC frame are ultimately modified at different levels.

For MP3,  many steganographic schemes have emerged\cite{bazyar2014recent,yang2017adaptive,zhang2017steganography}. Certain MP3 steganographic tools were published  early on, including MP3Stego\cite{petitcolas1998mp3stego}, UnderMP3Cover\cite{platt2004undermp3cover}, and MP3Stegz\cite{achmad2008mp3stegz}. Among them, MP3Stego\cite{petitcolas1998mp3stego}, which embeds the secret information by modifying the scale factor in the encoding procedure, is the most well-tested state-of-the-art MP3 steganography scheme.

With the development of  steganography schemes for AAC and MP3, many steganalysis schemes have followed \cite{ren2016steganalysis,yan2013steganalysis,ren2017steganalysis,yan2014detection,Yu2013Detecting,qiao2013mp3,kuriakose2015novel,Jin2014A,jin2017steganalysis}.
To detect  steganography schemes in the AAC Huffman coding domain, Ren\cite{ren2016steganalysis} proposed to extract the Markov transition probability of the neighbouring scale factor band's  (SFB's) codebook as the steganalysis feature to model the correlation of the neighbouring SFB's codebook. Calibration technology is used to improve the detection accuracy. To detect the steganography schemes in the AAC MDCT coefficient domain, Ren\cite{ren2017steganalysis} proposed to extract the Markov transition probabilities and joint probability densities of first-order and second-order differential residuals of the inter-frame and intra-frame MDCT coefficients as steganalysis features. The performance of the schemes\cite{ren2016steganalysis,ren2017steganalysis} is high for certain embedding domains but decreases for other embedding domains.

To detect MP3Stego\cite{petitcolas1998mp3stego}, which is one of the most well-tested MP3 steganography schemes,  many steganalysis schemes have been developed. Yan \cite{yan2013steganalysis} proposed to extract the difference between the quantization steps of MP3 neighbouring frames as the steganalysis features. Yan\cite{yan2014detection} proposed to extract the statistical inconsistencies in the statistical distribution of the number of bits in the audio bit pool before and after recompression as the steganalysis scheme. Yu\cite{Yu2013Detecting} used a recompression calibration technique and proposed a steganalysis scheme based on the statistical distribution features of Big\_value. Wang\cite{jin2017steganalysis} used Markov features to capture the correlations between quantized MDCT coefficients, which are destroyed by the MP3stego embedding procedure.
However, the steganalysis features of all the above schemes are extracted from a certain embedding domain; therefore, they achieve good performances for  steganographic schemes in an identical embedding domain.

With the successful application of Deep Neural Network (DNN) technology in various fields, DNNs have also been used in many steganalysis schemes \cite{Holub2013Random,Tan2014Stacked,qian2015deep,Xu2016Structural,qian2016learning,Xu2016Ensemble,chen2017audio,Sedighi2017Histogram,Yang2017Steganalysis,ye2017deep,wang2018cnn,boroumand2018deep,zeng2018large}. Among them,  two audio steganalysis schemes are based on DNNs. In \cite{chen2017audio}, a  convolutional neural network (CNN) is designed to detect audio steganography schemes in the time domain.
This work shows the effects of  CNNs in improving the performance of audio steganalysis schemes.
However, such schemes are effective at detecting the audio steganography scheme in the time domain but are  ineffective on  audio steganography schemes for AAC and MP3. In\cite{wang2018cnn}, to detect  MP3 steganography schemes in the Huffman coding domain, a steganalysis scheme based on CNNs was proposed. This scheme includes three main characteristics: the quantified MDCT matrix is used as the input of the CNN, a high-pass filter is used to enhance the difference introduced by the steganography, and many network optimization strategies are adopted. The experimental results show that the CNN-based scheme performs better than the hand-crafted steganalysis features. However, the same problem is found: the performance of the scheme is reduced to detect  other steganography schemes in different embedding domains.

According to  current research results in the field of steganalysis research, it can be concluded that DNN technology can significantly improve the detection accuracy of steganalysis\cite{chen2017audio,ye2017deep,wang2018cnn,boroumand2018deep}.
However, the key point is how to express the steganalysis data in a more rich and accurate manner and to design a more effective and suitable  DNN for steganalysis work. This paper attempts to address these points to improve the generality of steganalysis schemes for AAC and MP3. Although there are many embedding schemes for the audio compression parameters in AAC and MP3, such as the MDCT coefficients\cite{wang2011steganography}, the scale factor\cite{xu2009performance,wei2010controlling}, and Huffman coding\cite{wang2010steganography,zhu2010sign,zhu2011huffman}, all these schemes  change the MDCT values in different ways and ultimately change the time-frequency  correlations   of the audio signal. Therefore, if the steganalysis features are extracted from the time-frequency characteristics of the audio signal, the steganalysis scheme will be more generative and can be used to detect steganographic schemes that cause changes in the time-frequency characteristics. Almost all  steganographic schemes for  AAC and MP3 audio compression \cite{wang2011steganography,xu2009performance,wei2010controlling,wang2010steganography,zhu2010sign,zhu2011huffman,Luo2017Adaptive,
bazyar2014recent,yang2017adaptive,zhang2017steganography,petitcolas1998mp3stego,platt2004undermp3cover,achmad2008mp3stegz} belong to this category. Based on this idea, a steganalysis scheme called Spec-ResNet (Deep Neural Network of Spectrogram) is proposed in this paper. There are three main contributions provided by the proposed scheme.

\begin{itemize}
\item To extract the universal features introduced by the steganography schemes based on perceptual audio coding, the spectrogram, which is a visual representation of the spectrum of frequencies of an audio signal, is adopted as the input of the feature network. Because almost all steganographic schemes in the AAC and MP3 compressed domains produce changes in the spectrum energy relationships between time and different frequency bands,  the spectrogram can exactly reflect these relationships. 
\item To more effectively extract the classification features of steganography schemes, a deep residual network is used as the basic framework of the proposed scheme: Spec-ResNet. Because deep residual networks use deeper neural networks and do not suffer from the problem of gradient disappearance, they are suitable for extracting steganalysis features based on  weak signals.
\item To capture the steganalysis features more completely, the features extracted from multiple spectrograms of different window sizes are combined because the time-frequency relationships in the spectrogram are different under different window sizes. A simple SVM classifier is used to verify the performance improvement of the feature combination.
\end{itemize}

This paper is organized as follows. The related background knowledge is introduced in Section \uppercase\expandafter{\romannumeral2}. In Section \uppercase\expandafter{\romannumeral3}, the proposed steganalysis scheme, Spec-ResNet, is described in detail. The experiments results follow in Section \uppercase\expandafter{\romannumeral4}. Finally, the conclusion is drawn, and future work is discussed in Section  \uppercase\expandafter{\romannumeral5}.

\section{Preliminary}
\subsection{The principle of audio codecs}
\begin{figure}[!t]
\includegraphics[width=3.2in]{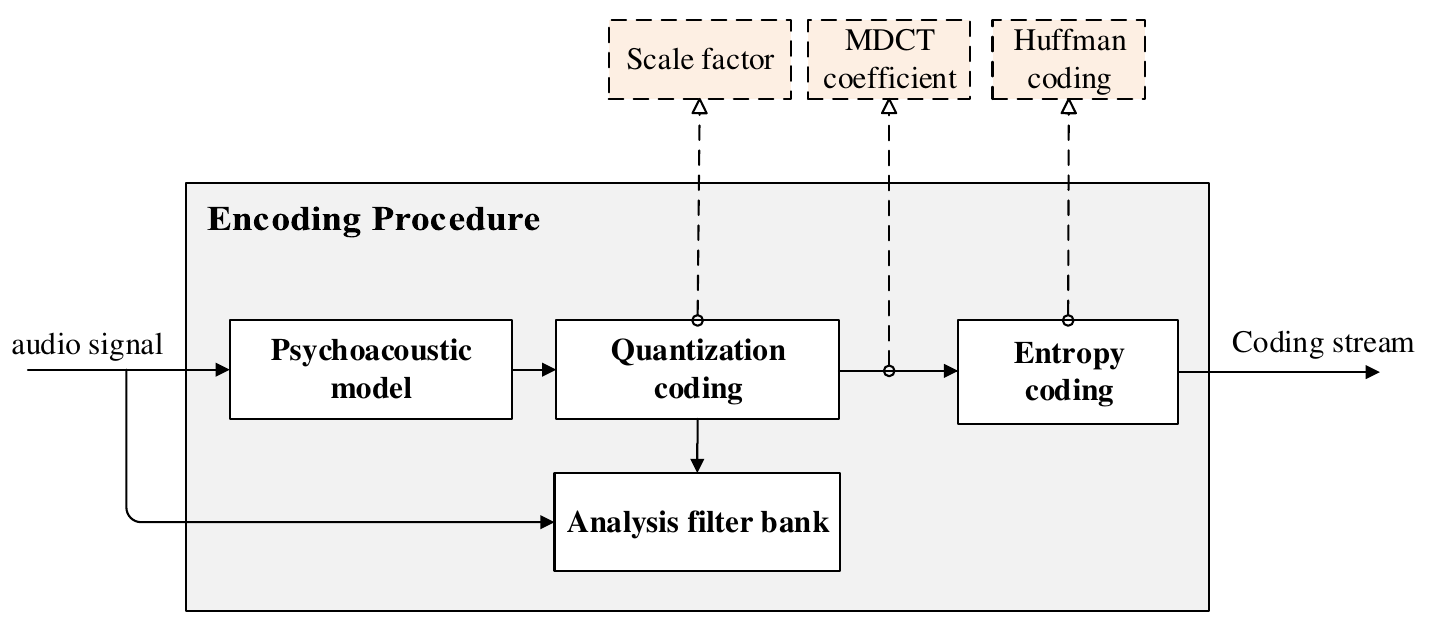}
\centering
\caption{The  perceptual audio coding procedure  and three main embedding domains.}
\label{fig_3_1}
\end{figure}

Both AAC and MP3 are under the framework of perceptual audio coding; therefore, their encoding principles, including the MDCT transform, psychoacoustic model, quantization and entropy coding, are similar. The advantages of AAC over MP3 include its higher sampling rates and number of channels supported, more efficient coding, and additional modules included to increase compression efficiency. The framework for the perceptual audio coding and the three main embedding domains are shown in Fig.\ref{fig_3_1}.
During the process of perceptual audio coding, the psychoacoustic model is used to first remove the part of the audio signal that  human hearing cannot perceive to reduce the data rate. Then, the input pulse signal is subjected to frame division or block division, and an analysis filter bank is used to perform time-frequency conversion. Each frequency domain sub-band is calculated according to a psychoacoustic model to obtain the corresponding masking threshold and maximum permissible distortion. Three types of cyclic quantization and entropy coding are performed on the transformed MDCT coefficient, and finally, the audio code stream of the entropy coding is output.

\subsection{The steganographic schemes of AAC and MP3}

Due to lossy quantization in AAC and MP3 coding, a slight modification of the quantized integer MDCT coefficients will not  obviously decrease the audio quality. Therefore, the embedding of a secret message can be implemented by fine tuning the value of the compression parameter. There are three main embedding domains: the MDCT coefficients, the scale factor and the Huffman coding parameters, as shown in Fig.\ref{fig_3_1}. The basic idea behind these steganography schemes is in modifying the compression parameters slightly to hide the secret information. Many AAC and MP3 steganography schemes have been developed; four such schemes from different embedding domains will be introduced in this section.

\subsubsection{LSB-EE}
In \cite{wang2010steganography}, the least significant bit (LSB) of the escape sequences in AAC Huffman coding are modified with matrix encoding to embed secret information. In the AAC codec, the escape sequence is a special codebook used to encode MDCT coefficients with values of greater than 16. The modification of the LSB of the escape sequence produces minimal negative effects on audio quality. In AAC compressed data, MDCT coefficients with values of greater than 16 account for  5\%-6\% of all coefficients, and they are mainly of low frequencies; therefore,  changes in this scheme to the AAC audio signal are almost always in the low-frequency part of the signal.

\subsubsection{MIN}
In\cite{wang2011steganography}, Wang proposed a method for implementing steganography in the small-value region of the MDCT coefficients because the energy of the audio is mainly concentrated in the low- and middle-frequency portions, the MDCT coefficients of which are quantized with a smaller quantization step size, and a large quantization step size is used for the high-frequency portion to reduce the distortion as much as possible. For the high-frequency portion, the quantization range is larger, thereby requiring fewer coding bits. The amplitude of the quantized coefficient is basically \{-1, 0, 1\}, which is called the small-value region of the MDCT coefficient. In the AAC quantized MDCT coefficient encoding stage, the coefficients of the small value region are generally encoded by codebook 1 and codebook 2, and the 4 quantized MDCT coefficients are grouped into one index value: $index$. The calculation  is given in (\ref{equ:2}).
 \begin{equation}\label{equ:2}
index=\sum_{i=0}^3{3^i}\times q\left[ i \right] +40,q\left[ i \right] \in \left\{ -1,0,1 \right\}
\end{equation}
The codeword is searched among the codebooks according to $index$. To further reduce the distortion in AAC audio, the steganographic algorithm uses the method of modifying the last MDCT coefficient, $q[3]$, and calculating a new $index$ to replace the original $index$.
In AAC compressed data, the small MDCT coefficients are mainly in the low- and mid-frequency portions; therefore, the changes induced by this scheme to the AAC audio signal are generally in the low- and mid-frequency portions of the signal.

\subsubsection{SIGN}
In\cite{zhu2010sign}, Zhu proposed to embed a message in the Huffman encoding process of the AAC coding stream. The MDCT coefficients are encoded by a plurality of codebooks. According to the number of bits after encoding, the encoder selects an optimal codebook to encode 2 or 4 MDCT coefficients as a Huffman code. If the codebooks selected have sign bits, the sign bits of the non-zero MDCT coefficients are attached to the codeword. The coding stream structure corresponding to the codebook with sign bits is shown in Table \ref{tab1_1} and Table \ref{tab1_2}.  In codebooks 1 and 2 or codebooks 5 and 6, $Huffman\_code$ represents 4 or 2 MDCT coefficients, while $sign\_w,\ sign\_x,\ sign\_y,\ sign\_z$ are the sign bits of MDCT coefficients $w,\ x,\ y,\ z$, respectively. $sign=0$ represents that the MDCT coefficient is negative,  and $sign=1$ represents that the MDCT coefficient is positive.
 \begin{table}[htbp]
   \centering
   \caption{AAC coding stream structure of codebooks 1 \& 2.}
   \setlength{\tabcolsep}{1mm}{
     \begin{tabular}{ccccccc}
     \toprule
     бн & $Huffman\_code$ & $sign\_w$ & $sign\_x$ & $sign\_y$ & $sign\_z$ & бн \\
     \bottomrule
     \end{tabular}}%
   \label{tab1_1}%
 \end{table}%

 \begin{table}[htbp]
   \centering
   \caption{AAC coding stream structure of codebooks 5 \& 6.}
     \begin{tabular}{ccccc}
     \toprule
     бн & $Huffman\_code$ & $sign\_x$ & $sign\_y$ & бн \\
     \bottomrule
     \end{tabular}%
   \label{tab1_2}%
 \end{table}%
 Given an appropriate threshold T, SIGN modifies the sign bit of the MDCT coefficients ranging from $-T$ to $T$ accordingly to implement audio steganography.
 This subtle modification will not change any parameters in the AAC encoding procedure except for the sign bit of a specific MDCT coefficient; therefore, it produces minimal distortion in the audio quality.


\subsubsection{MP3Stego}
In\cite{petitcolas1998mp3stego}, MP3Stego was developed as a public tool to hide information in MP3 files during the MP3 compression process. There are two nested loops used to quantify the MDCT coefficients in the MP3 encoder: an inner loop and an outer loop. The outer loop is used to verify the distortion of the audio signal to guarantee satisfactory audio quality. The inner loop is used to select a suitable scalar factor to qualify the MDCT coefficients with the available number of bits. The hiding operation used by MP3Stego is performed in the inner loop. The \emph{part2-3-length} variable, which records the number of main-data bits, is controlled by  selecting  the scalar factor to make the LSB of the \emph{part2-3-length} variable equal to the secret bit.

All  four steganographic schemes modify the compression parameters to hide information. Although the embedding domain is different, the final result of these schemes is changes in the MDCT coefficients of the encoded audio signals in different frequency ranges. Because the above four schemes cover almost the entire frequency range of audio signals, they are chosen as the target steganography schemes to verify the generalizability of the proposed steganalysis scheme.

\subsection{Spectrogram}
\begin{figure}[!t]
\includegraphics[width=3.0in]{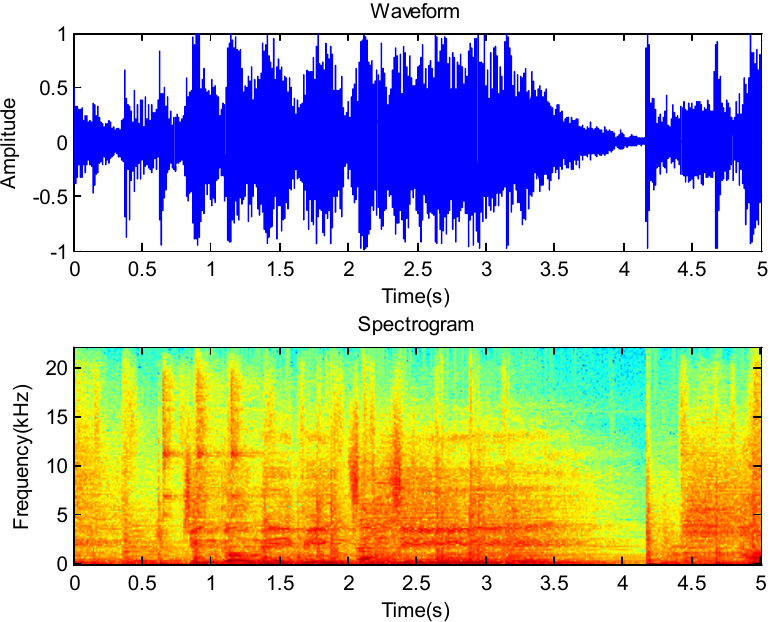}
\centering
\caption{Waveform and spectrogram of an audio segment.}
\label{fig_2}
\end{figure}

A spectrogram is a basic visual representation of the spectrum of frequencies of audio signals. A spectogram can show the energy amplitude information of different frequency bands over time,  and it contains abundant time-frequency information about the audio signal. Therefore, spectrograms are a good research object for general audio steganalysis when attempting to capture more relative features introduced by different audio steganography schemes. Fig. \ref{fig_2} is the spectrogram of an audio segment together with its waveform.

In the spectrogram, the horizontal axis represents the time domain  of the audio signal, and the vertical axis represents the frequency domain; the third dimension, which is represented by the intensity of the colour of each point in the image, indicates the amplitude of the frequency at that time. To plot the spectrogram of the audio signal, the signal is divided into many frames under a given window size. The formant of the spectrogram can effectively express the acoustic parameters, and the spectrogram's local characteristics can accurately represent additive noise; therefore, it is an effective candidate for capturing the traces left by embedding operations.

\section{The architecture of Spec-ResNet}
\begin{figure*}[htbp]
\includegraphics[width=5in]{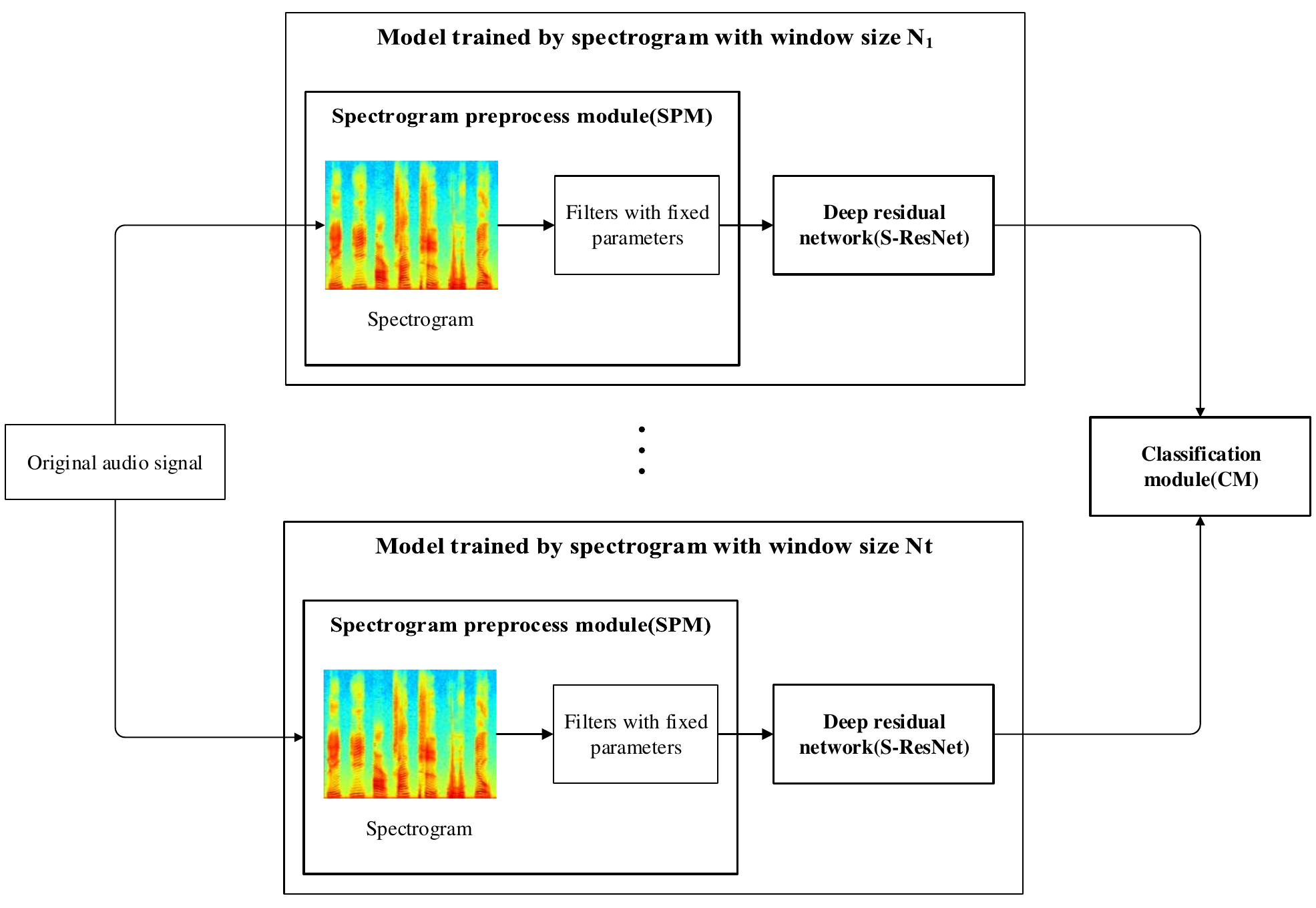}
\centering
\caption{Architecture of the proposed steganalysis scheme: Spec-ResNet.}
\label{fig_4}
\end{figure*}

Based on the above analysis, a universal audio steganalysis scheme, Spec-ResNet, is proposed in this paper. The spectrogram of an audio signal is adopted as the steganalysis object, a deep residual network is used to extract the steganalysis features, and several spectrograms under different window sizes are combined  to extract more rich features.

The main architecture of the Spec-ResNet scheme, which consists of three main components, i.e., the Spectrogram Preprocess Module (SPM), the Deep Residual Steganalysis Network (S-ResNet) and the Classification Module (CM), is shown in Fig. \ref{fig_4}. SPM generates the spectrogram for the given window size of the input audio signal and uses filters to preprocess the spectrogram. S-ResNet is the steganalysis feature network framework based on the  deep residual network architecture and is used to extract the steganalysis features of the filtered spectrogram. The CM is a traditional classification process, which fuses the features from multiple spectrograms under different window sizes. The details of each module are described below.

\subsection{Spectrogram Preprocess Module}
\begin{figure}[!t]
    \includegraphics[width=3.5in]{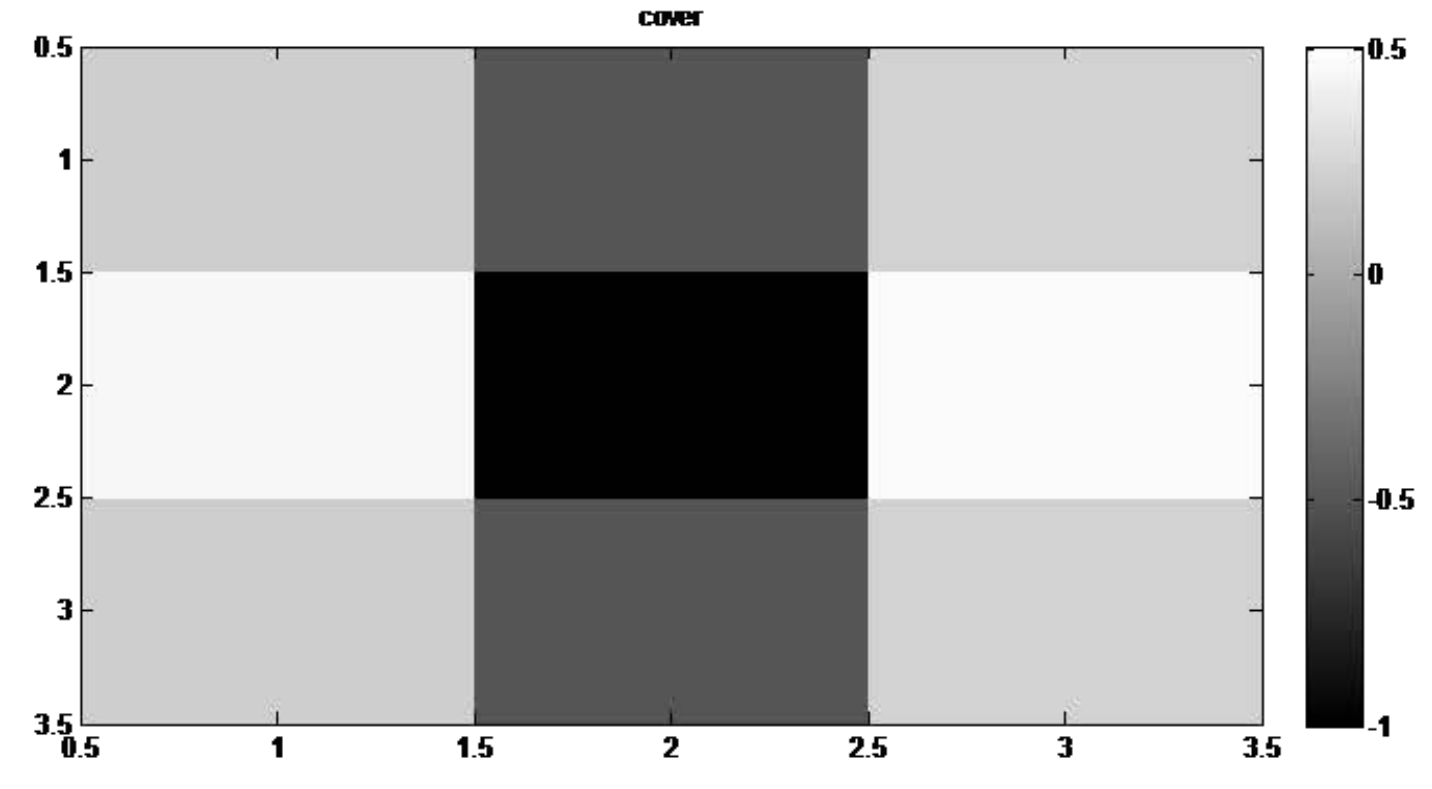}
    \centering
    \caption{The results  of the linear regression. The depth of the colour or  the absolute value of logistic coefficients represent the strength of the correlation.}
    \label{fig_5}
\end{figure}

In this module, the trained or test audio signals are represented as a spectrogram under a given window size. Audio signals are highly correlated in both the time and frequency domains; therefore, the spectrogram is a highly effective representation for analysing the correlation changes in the time and frequency domain  introduced by the steganography operation.
The spectrogram of the audio signal contains three type of signals: the audio content signal, the noise of the real audio signal and the noise introduced by embedding operations. As in the  image steganalysis tasks, the noise introduced by steganography is weak; therefore, the  audio signal noise should be amplified to improve the signal-to-noise ratio (SNR) of the steganography noise.
In\cite{fridrich2012rich}, Fridrich \emph{et al.} proposed an image steganalysis scheme based on the spatial domain rich model feature (SRM), in which multiple filters are introduced to capture the noise distribution characteristics from the tested image in different directions and dimensions. Inspired by this work, the correlations between intra-frame and inter-frame information in the spectrograms of 100 cover audio samples are analysed by linear regression, and multiple filters are designed to preprocess the spectrogram.
The audio signals have a duration of 2 s at 32 kHz, are AAC files, and are decoded as WAV files. The spectrogram  is an $n\times m$ spectrogram, where $n$ is half the window size due to the symmetry in the Fourier transform and $m$ is the total number of frames in one piece of audio with 50\% overlap (e.g., if the window size $N=1024$, the dimension of the spectrogram is $512\times 128$). The multiple linear regression model is shown in (\ref{equ:4}). $y$ is the dependent variable, $x_t,t=1,2,...,s$ is the independent variable, and $\beta _0$ is the bias. For the $3\times 3$ filter,  $s$ is 8, and the correlation between the matrix centre point and the 8 neighbouring signal points is presented in Fig. \ref{fig_5}.
\begin{equation}\label{equ:4}
y=\beta _0+\beta _1\times x_1+...+\beta _t\times x_t+...+\beta _{s}\times x_{s}
\end{equation}

In Fig.\ref{fig_5},  the correlations between the neighbouring points in the horizontal and vertical directions are relatively strong, especially in the vertical direction. In the spectrogram, the vertical direction correlation is the relevance of the neighbouring frequency band in the same frame, and the horizontal direction describes the relation between  successive frames in the same frequency band. Based on the above analysis, a set of filters from different directions are chosen as the first convolutional layer of the deep residual network to extract the energy difference introduced by the steganographic noise between the cover and stego samples.

\begin{figure}[!t]
\includegraphics[width=2.5in]{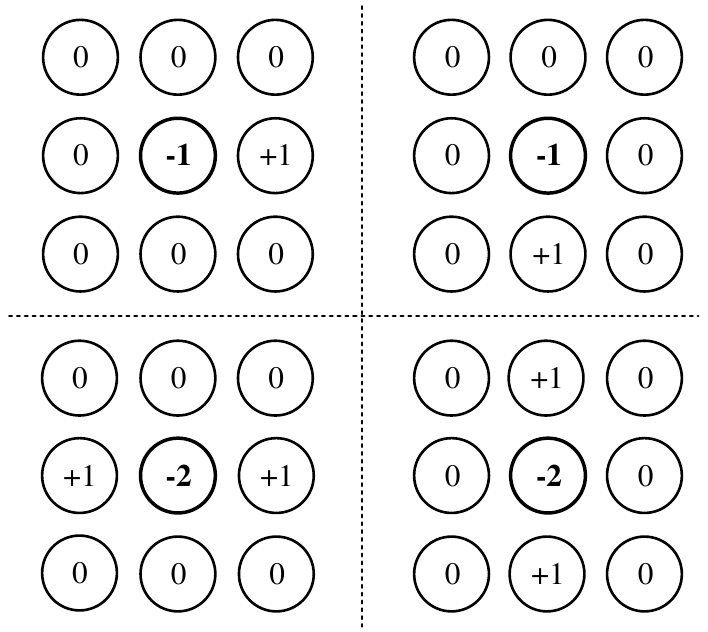}
\centering.
\caption{The filters based on the  intra-frame and inter-frame correlations.}
\label{fig_7}
\end{figure}

In this work, 4 fixed-parameter filters are chosen to preprocess the spectrogram. The size of the filters is $3\times 3$, and blanks are filled with zeroes, as shown in Fig. \ref{fig_7}. As future work, the filter parameters can be trained by the network to improve the performance of the proposed scheme.

\subsection{Deep Residual Steganalysis Network}
\begin{figure*}[htbp]
\includegraphics[width=5in]{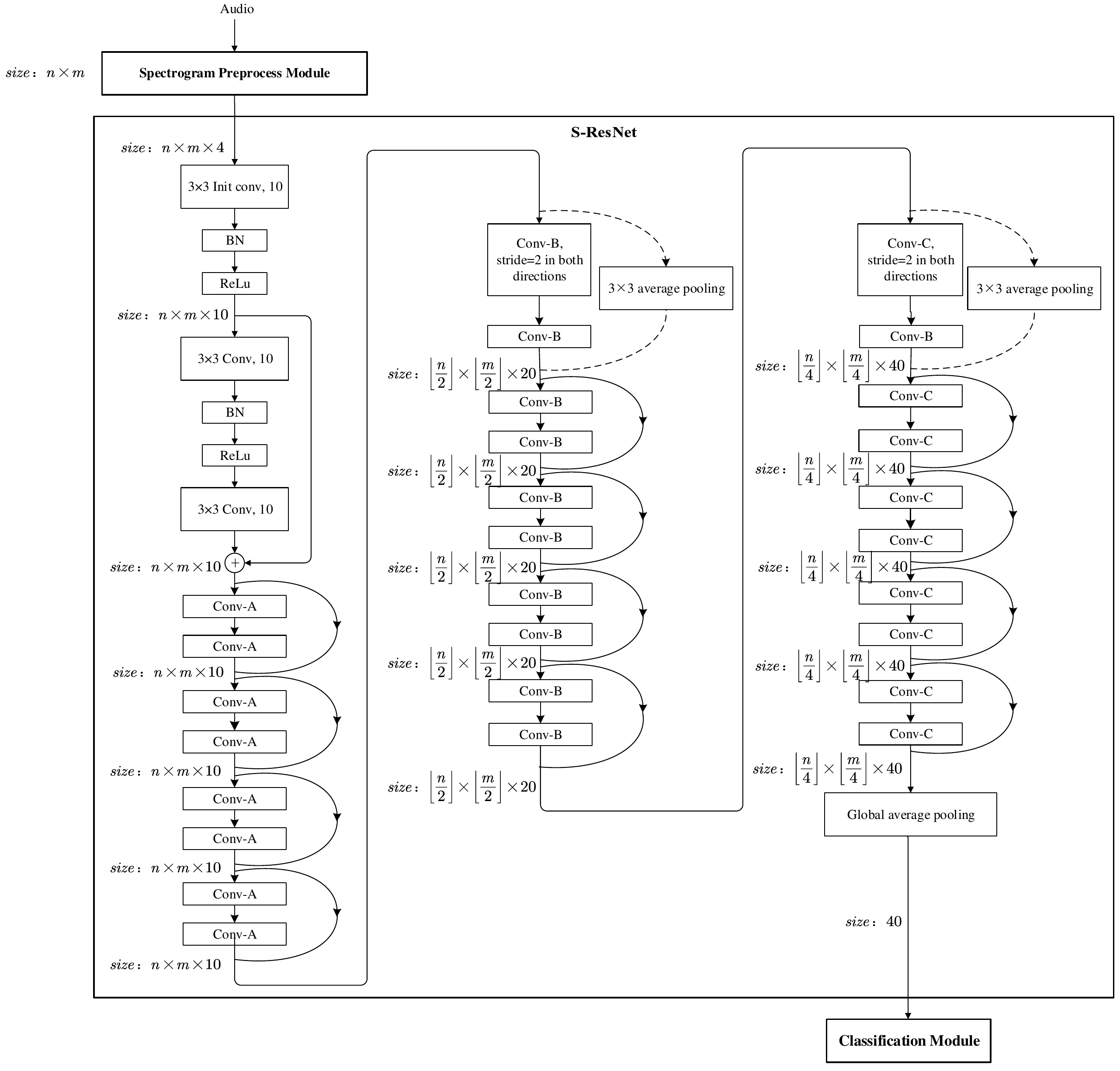}
\centering.
\caption{The structure of the S-ResNet deep residual steganalysis network.}
\label{fig_9}
\end{figure*}

The spectrogram of the audio signal is filtered using the four filters. Then, it is processed by the initial convolutional layer with 10 output channels to extract the potential correlations in the feature matrix. S-ResNet consists of 30 layers, and there is a shortcut for each pair of convolutional layers. A global average pooling layer is subsequently applied. The  S-ResNet architecture  is shown in Fig. \ref{fig_9}, and details about  S-ResNet are as follows.

\subsubsection{Convolutional layers}
\begin{figure}[t]
	\includegraphics[width=2.0in]{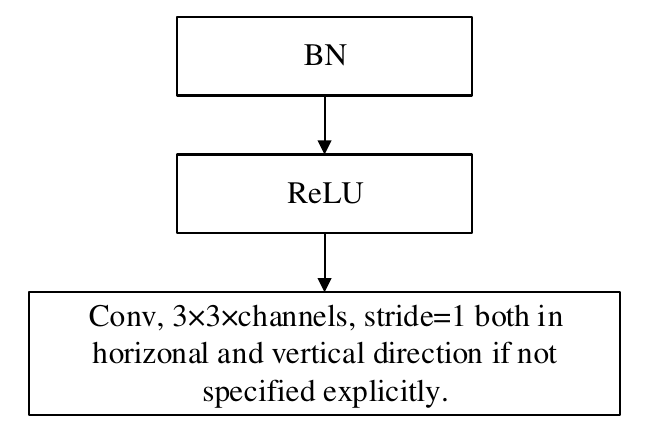}
	\centering.
   	 \caption{ The structure of the convolutional layer. Conv-A has 10 channels, Conv-B has 20 channels, and Conv-C has 40 channels.}
    \label{fig_8}
\end{figure}

The convolutional layers in this paper are a $3 \times 3$ kernel, regardless of whether the parameters of the filters are fixed. The filters introduced in SPM are applied to weaken the influence of the audio content in the spectrogram. The remaining convolutional layers, which include the initial convolutional layer after SPM, are used to capture the potential correlations inherent in the feature matrix. Based on the number of channels, the convolutional layers in the residual network can be classified into three types: Conv-A, Conv-B, and Conv-C, as shown in Fig. \ref{fig_8}. There are 10, 20, and 40 channels in Conv-A, Conv-B, and Conv-C, respectively, and the stride is 1 in both  the horizontal and vertical directions unless otherwise stated. Moreover, there are 5 residual units in each group, which means 10 convolutional layers.
\subsubsection{Residual unit}
Generally, for  Convolutional Neural Networks (CNNs), more convolutional layers means that the features extracted from different dimensions are richer and will provide higher classification accuracy. However, once the CNN reaches a certain depth,  increasing the number of neural network layers may lead to a degradation of the performance due to the vanishing gradient problem. At this time, the training and test accuracies will  decrease; therefore, it would be difficult to train an effective neural network, and the learning ability of the deep neural network will be limited.
\begin{figure}[!t]
\includegraphics[width=2.5in]{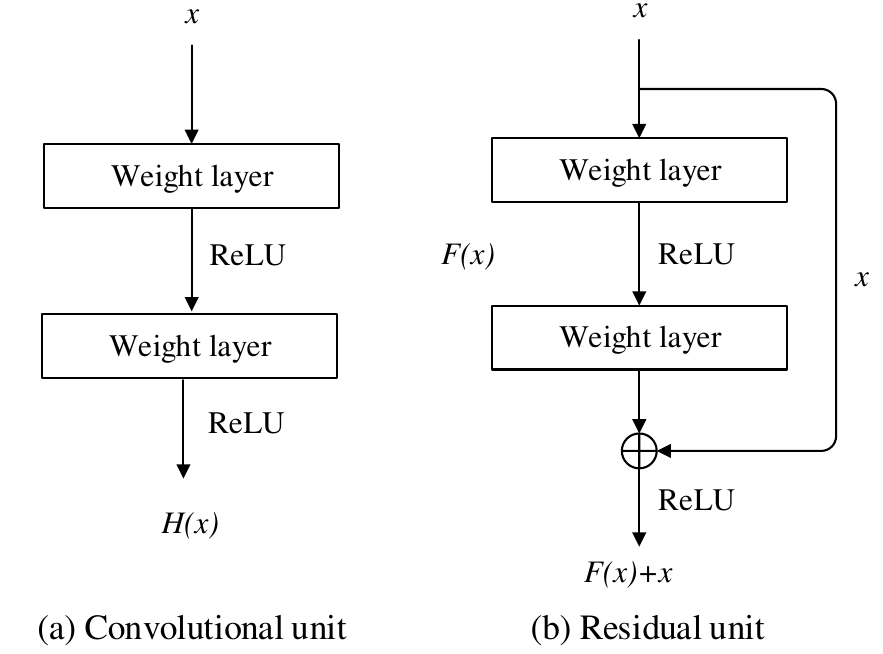}
\centering.
\caption{The structure of the convolutional unit and the residual unit.}
\label{fig_6}
\end{figure}

Compared with  classical CNNs, the residual network\cite{He2015Deep} is capable of integrating more convolutional layers without suffering a decrease in network performance. The model learning can be simplified to approximate a mapping function learning, as described in Fig.\ref{fig_6}. $H\left( x \right) =x$ is the hypothesis model of the cascaded convolutional layers. The objective of the CNN is in attempting to find a mapping function $H\left( x \right)$ directly. However, a residual network is different due to the identity function $x$; a residual network attempts to find a residual function $F\left( x \right) =H\left( x \right) -x$ instead. In this way, the residual network is transformed to approximate an ideal hypothesis by learning the residual $F\left( x \right) =0$, which is easier than approximating $H\left( x \right)$ directly. Obviously, the convolutional output is more sensitive to  changes in the input than previously with the identity function $x$ introduced.

In Fig.\ref{fig_9}, the arc across each pair of convolutional layers represents the shortcut. The real arc indicates that $x$ has the same size as $F(x)$, and the two tensors can be added together directly. A dashed arc between Conv-A and Conv-B or Conv-B and Conv-C indicates that they have different sizes, and $x$ requires downsampling before the add operation.
\subsubsection{Batch normalization layer}
Batch normalization was proposed in \cite{Ioffe2015Batch} to accelerate deep network training. The distributions of the inputs before each hidden layer vary in each training iteration. Batch normalization performs normalization for each training iteration instead of setting lower learning rates at first or restricting parameter initialization, which requires more time to train the model. Batch normalization is applied before each new convolutional layer in Fig.\ref{fig_9}.
\subsubsection{Activation function}
In addition to batch normalization, an activation function is  applied after each batch normalization in S-ResNet. Nonlinear links between convolutional layers provide better expression ability than do linear model; therefore, the Rectified Linear Unit  (ReLU) is adopted as the activation function in this paper.
\subsubsection{Pooling layers}
To reduce the computational complexity, a max pooling layer or average pooling layer is used in the deep network to reduce the feature dimensions. A $3 \times 3$ average pooling layer is applied to S-ResNet between Conv-A and Conv-B and between Conv-B and Conv-C with $stride = 2$ in both the horizontal and vertical directions. There are no pooling layers in the shortcut across Conv-A, Conv-B, or Conv-C to keep the dimensions of $x$ and $F(x)$ consistent. A global average pooling is applied immediately before the classification module (CM) to obtain a 40-dimensional classification feature.
\subsection{Classification module}
\begin{figure}[htbp]
\includegraphics[width=3.0in]{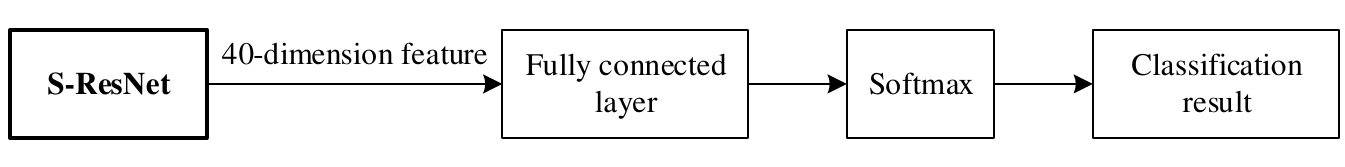}
\centering.
\caption{The classification module for Spec-ResNet of the spectrogram under the given window size.}
\label{fig_10}
\end{figure}

In other CNN-based\cite{chen2017audio,wang2018cnn} and RNN-based\cite{Lin2018RNN} steganalysis schemes, the classification module in the network is always a fully connected layer with a 2-way softmax layer, as shown in Fig.\ref{fig_10}, for obtaining the probabilities of 2 class labels. With a given threshold ranging from 0 to 1 (default of 0.5), the tested samples are judged as cover or stego.

Under the short-time Fourier transform, different spectrograms with different scales will be generated when different window sizes are used. When the window size $N$ is small, a broad-band spectrogram will be generated, the time bandwidth is narrow, and the corresponding frequency bandwidth is broad. A narrow-band spectrogram will produce the opposite characteristics. A broad-band spectrogram can well reflect the time-varying characteristics of the spectrum; however, the frequency resolution is low and does not accurately reflect the texture characteristics of the sound. The frequency resolution of the narrow-band spectrogram is too small to properly identify the position of the formant.

\begin{figure}[htbp]
\includegraphics[width=3in]{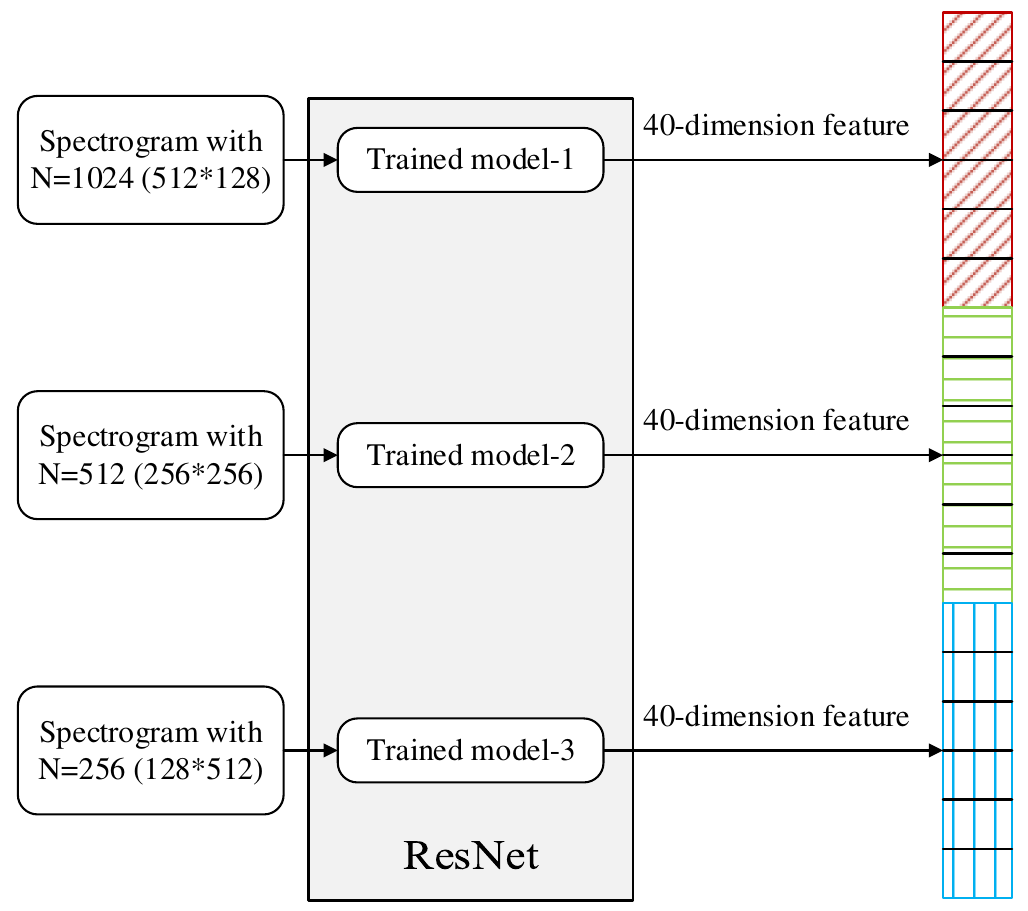}
\centering.
\caption{The classification module for Spec-ResNet with multiple spectrograms under different window sizes.}
\label{fig_11}
\end{figure}

In the spectrogram feature map, the audio signal characteristics are different under different window sizes; therefore, it is very important to select a  suitable window size for steganalysis. In this paper, a spectrogram with different window sizes is considered. For a given audio sample, its spectrogram has different dimensions with different $N$; however, the classification feature's dimension after global average pooling is constant. In our experiment, three spectrograms with different window sizes are chosen to increase the diversity of features. A total of three 40-dimensional features are cascaded to a 120-dimensional feature as the final steganalysis feature. Fig. \ref{fig_11} presents the classification module used in this paper, and the window sizes N are 1024, 512, and 256. LibSVM\cite{chang2011libsvm} is introduced to train the ultimate classification model based on the 120-dimensional feature originating from the audio samples. In  detection stage, the 120-dimensional feature extracted from the test audio samples is used to judge whether a piece of audio is cover or stego audio.
\section{Experiments}
To evaluate the performance of the proposed  Spec-ResNet scheme, three  experiments have been performed. The purpose of the first experiment is to evaluate the detection accuracy of the proposed scheme for the three AAC steganography schemes with different embedding domains. Spectrograms with three different window sizes are used to evaluate the influence on the detection performance. The second experiment is implemented to compare Spec-ResNet with work on other steganalysis schemes. The third experiment is performed to detect  MP3Stego, which uses the MP3 codec and whereby the embedding domain is the scalar factor.

\subsection{Experimental setup}
\subsubsection{Dataset}

To the best of the authors' knowledge, there are no  public audio datasets for AAC or MP3 steganalysis; therefore, we constructed an in-house audio dataset. A total of 186 recordings of music, including different musical styles, such as jazz, rock, country, pop; different languages (mainly Chinese and English); and different singer genders, were downloaded from the Internet. The music samples were cut into clips to construct the \emph{AAC audio dataset} and the \emph{MP3 audio dataset}. All the audio datasets in the experiments described below. The project is based on Tensorflow 1.0.0 + CUDA 8.0.61 + CuDNN 5.1.10 + Python 2.7.15. In addition, all parameters are trained with the Adam optimizer: the mini-batch size is 32, which means that 16 pairs of cover and stego audio samples are used; the learning rate decay is 0.9; and the weight decay is $2\times 10^{-4}$.

\paragraph{AAC audio dataset}
The 10,000 16-kHz, 16-bit, 2-s long audio samples are used to construct 4 different audio sample sets, including one cover sample set (\emph{CDB\_AAC}) and three stego sample sets (\emph{SDB\_AAC\_LSB}, \emph{SDB\_AAC\_MIN}, \emph{SDB\_AAC\_SIGN}). The main construction of the four datasets is as follows.

\textbf{\emph{CDB\_AAC}}: The audio samples in WAV format are encoded using the open-source audio encoder FAAC\cite{encoder-decoder} at 32 kbps and then decoded to WAV using FAAD2. The decoded audio samples are  32 kHz with 2 channels, and the total number of cover samples in \emph{CDB\_AAC} is 10,000.

\textbf{\emph{SDB\_AAC\_LSB}}: The 10,000 audio samples (.m4a format) are generated by the LSB\_EE steganographic scheme \cite{wang2010steganography}, and the secret message is embedded with  relative embedding rates (EBRs) of 10\%, 20\%, 30\%, 50\%, and 100\%. EBR is the ratio of the embedded message length to the maximum embedded message length. The total number of stego samples (.wav format) in \emph{SDB\_AAC\_LSB} is $10,000\times 5=50,000$ after the compressed audio bitstream is decoded using FAAD2.

\textbf{\emph{SDB\_AAC\_MIN}}: The 10,000 audio samples (.m4a format) are generated by the MIN steganographic scheme \cite{wang2011steganography}, and the secret message is embedded with EBRs of 10\%, 20\%, 30\%, 50\%, and 100\%. The total number of stego samples (.wav format) in \emph{SDB\_AAC\_MIN} is $10,000\times 5=50,000$ after the compressed audio samples are decoded using FAAD2.

\textbf{\emph{SDB\_AAC\_SIGN}}: The 10,000 audio samples (.m4a format) are generated by the SIGN steganographic scheme \cite{zhu2010sign}, and the secret information is embedded with  EBRs of 10\%, 20\%, 30\%, 50\%, and 100\%. The total number of stego samples (.wav format) in \emph{SDB\_AAC\_SIGN} is $10,000\times 5=50,000$ after the compressed audio samples are decoded with FAAD2.

\paragraph{MP3 audio datasets}
A total of 9,000 16-bit, 44.1 kHz, 5-s-long audio sample segments are used to construct 2 different audio datasets, including a cover sample set(\emph{CDB\_MP3}) and a stego sample set(\emph{SDB\_MP3}). The main contents of the two datasets are as follows.

\textbf{\emph{CDB\_MP3}}: The audio samples in the WAV file format are encoded at 128 kbps and decoded using a publicly available MP3 codec. The decoded audio samples (.wav format) are 44.1 kHz samples, and the total number of cover samples in \emph{CDB\_MP3} is 9,000.

\textbf{\emph{SDB\_MP3}}: The audio samples in the WAV file are encoded using MP3Stego\cite{petitcolas1998mp3stego}. Because the maximum embedding capacity of MP3Stego is approximately 2 bits per frame and because a 5 s audio sample into which a 3-byte message is embedded has 191 frames,  the EBR is nearly 6\%. In addition, the total number of stego samples in \emph{SDB\_MP3} is 9,000.

\subsubsection{Metrics}

To evaluate the detection accuracy of the proposed scheme and to compare schemes, the TPR (True Positive Rate), TNR (True Negative Rate) and ACC (Classification Accuracy) are adopted as metrics. TPR indicates the proportion of stego samples that are detected as stego, TNR means the proportion  of cover samples that are  detected as cover, and ACC is the detection accuracy for all  cover and stego samples.

\subsection{Experiments and Results}
\subsubsection{Experiment \uppercase\expandafter{\romannumeral1}}
This experiment is performed to evaluate the detection accuracy of the proposed scheme. Three AAC steganography schemes with different embedding domains are detected.
Spectrograms with three different window sizes $N_{win}=256,\ 512,\ or\ 1,024$  are extracted from 6,000 audio samples in \emph{CDB\_AAC}, \emph{SDB\_AAC\_LSB}, \emph{SDB\_AAC\_MIN}, and \emph{SDB\_AAC\_SIGN}. In the training procedure, the spectrograms extracted from 3,000 audio samples in \emph{CDB\_AAC} are used as the cover input signal, and the spectrograms of 3,000 audio samples with 0.3 EBR in \emph{SDB\_AAC\_LSB}, \emph{SDB\_AAC\_MIN} and \emph{SDB\_AAC\_SIGN} are selected as the stego input signal. For a given window size, the training set contains 6,000 feature matrices, as described above (for $N=256$, the dimensions of the spectrogram feature matrix are $128\times 512$; for $N=512$, the dimensions of the spectrogram feature matrix are $256\times 256$; and for $N=1024$, the dimensions of the spectrogram feature matrix are $512\times 128$). However, the architecture of the test dataset is different. For the cover samples, the spectrogram of the remaining 3,000 audio samples in \emph{CDB\_AAC} is selected, and spectrogram of the corresponding 3,000 audio samples generated by the different schemes and with different EBR (200 samples in each algorithm for a given EBR) is selected as the stego samples. Three models are trained based on the spectrogram with three different window sizes.

The results of experiment \uppercase\expandafter{\romannumeral1} are shown in Table \ref{tab_1}. $N=256,\ 512,\ or\ 1024$ represents the spectrogram window size. \textbf{EA} is the embedding scheme, and \textbf{ACC} is the average detection accuracy of the trained model toward cover or stego samples generated by a given steganographic algorithm with different EBR. \textbf{AVERAGE} represents the average detection accuracy for the cover or stego samples under a given EBR for multiple schemes. The results show that the performances of the classification model trained by a spectrogram with different window sizes are different. The average accuracy for cover samples under different models is always greater than 83.89\%. When $N=512$, the accuracy for the cover samples can is up to 93.47\%, and the average detection accuracy for the stego samples with 10\% EBR is greater than 82.6\%. When $N=1024$, the accuracy reaches 89.05\%. This shows that the proposed  Spec-ResNet scheme can be used to detect multiple steganographic schemes in different embedding domains and achieve good performance.

\begin{table*}[htbp]
  \centering
  \caption{The performance of the Spec-ResNet steganalysis scheme  using spectrograms with different window sizes}
  \begin{spacing}{1.35}
  \setlength{\tabcolsep}{3mm}{
    \begin{tabular}{rcccccccc}
    \toprule
    \multicolumn{1}{c}{\multirow{1}[6]{*}{\textbf{Window Size}}} & \multirow{1}[6]{*}{\textbf{EA}} & \textbf{TNR} & \multicolumn{5}{c}{\textbf{TPR}} &
    \multirow{1}[6]{*}{\textbf{ACC}} \\
    \multicolumn{1}{c}{} &   & \textbf{Cover} & \textbf{10\%} & \textbf{20\%} & \textbf{30\%} & \textbf{50\%} & \textbf{100\%} \\
    \midrule
    \multicolumn{1}{c}{} & LSB-EE\cite{wang2010steganography} & \multirow{2}[6]{*}{0.9044} & 0.9433 & 0.9570 & 0.9675 & 0.9796 & 0.9898 & 0.9359\\
    \multicolumn{1}{c}{$N$=1024} & MIN\cite{wang2011steganography} &   & 0.9329 & 0.9541 & 0.9571 & 0.965 & 0.9646 & 0.9296 \\
      & SIGN\cite{zhu2010sign} &   & 0.7954 & 0.8388 & 0.8592 & 0.8921 & 0.925 & 0.8833\\ \hline
    \multicolumn{2}{r}{\textbf{AVERAGE} \quad} & \textbf{0.9044} & \textbf{0.8905} & \textbf{0.9166} & \textbf{0.9279} & \textbf{0.9456} & \textbf{0.9598} &  \textbf{0.9163}\\ \hline
    \multicolumn{1}{c}{} & LSB-EE\cite{wang2010steganography} & \multirow{2}[6]{*}{0.9347} & 0.9392 & 0.9679 & 0.9758 & 0.9923 & 0.9979 & 0.9547 \\
    \multicolumn{1}{c}{$N$=512} & MIN\cite{wang2011steganography} &   & 0.9413 & 0.9617 & 0.9738 & 0.9708 & 0.975 & 0.9496 \\
      & SIGN\cite{zhu2010sign} &   & 0.7704 & 0.8054 & 0.8342 & 0.8671 & 0.9208 & 0.8871 \\ \hline
    \multicolumn{2}{r}{\textbf{AVERAGE} \quad} & \textbf{0.9347} & \textbf{0.8836} & \textbf{0.9117} & \textbf{0.9279} & \textbf{0.9434} & \textbf{0.9646} & \textbf{0.9305} \\ \hline
    \multicolumn{1}{c}{} & LSB-EE\cite{wang2010steganography} & \multirow{2}[6]{*}{0.8389} & 0.8871 & 0.9179 & 0.9446 & 0.9292 & 0.9438 & 0.8817 \\
    \multicolumn{1}{c}{$N$=256} & MIN\cite{wang2011steganography} &   & 0.9329 & 0.9471 & 0.9613 & 0.9692 & 0.9833 & 0.8988 \\
      & SIGN\cite{zhu2010sign} &   & 0.6579 & 0.7317 & 0.7821 & 0.8063 & 0.875 & 0.8048 \\ \hline
    \multicolumn{2}{r}{\textbf{AVERAGE} \quad} & \textbf{0.8389} & \textbf{0.8260} & \textbf{0.8656} & \textbf{0.896} & \textbf{0.9016} & \textbf{0.934} & \textbf{0.8618} \\
    \bottomrule
    \end{tabular}}
    \end{spacing}%
  \label{tab_1}%
\end{table*}%

\subsubsection{Experiment \uppercase\expandafter{\romannumeral2}}
This experiment is performed to compare the detection accuracy of the proposed scheme with other steganalysis schemes.
Although the feature matrix of a spectrogram with different window sizes has a different shape, the length of the feature following the global average pooling layer is constant, i.e., 40. For a spectrogram of a given window size, a classification model is trained based on this window size and will be saved thereafter. Then, three 40-dimensional features originating from three trained classification models are merged into a 120-dimensional feature. SVM is used to train the ultimate classification model, therein considering spectrogram features with different scales. In particular, the architecture of the training  and test sets is the same as that in \emph{Experiment \uppercase\expandafter{\romannumeral1}}.

The measure the performance of Spec-ResNet, it is compared with Ren's scheme\cite{ren2017steganalysis}, Luo's scheme\cite{chen2017audio}, and Zhao's scheme\cite{wang2018cnn}. In Ren's scheme\cite{ren2017steganalysis}, a classical rich model feature is introduced to fully analyse the Markov transition probabilities and joint probability densities of the first-order differentials and second-order differential residuals of inter-frame and intra-frame quantified MDCT coefficients. In Luo's scheme\cite{chen2017audio}, the classifier is trained with seven convolutional neural networks based on the original audio signal in the time domain.
In\cite{wang2018cnn}, a steganalysis scheme based on a CNN of the quantified MDCT matrix is proposed to detect the MP3 steganography schemes in the Huffman coding domain. In this experiment, the scheme in \cite{wang2018cnn} is realized in the AAC codec, and the CNN is replaced with S-ResNet; this will improve the performance of the scheme in\cite{wang2018cnn}. Audio samples with a duration of 2 s are adopted for performance testing in this experiment, and 8,000 audio samples, including 4,000 cover samples and 4,000 stego samples, are used for training. The remaining 12,000 audio samples are used for testing.

\begin{figure*}
  \centering
  \subfloat[]{
\begin{minipage}[t]{0.3\linewidth}
\centering
\includegraphics[width=2.2in]{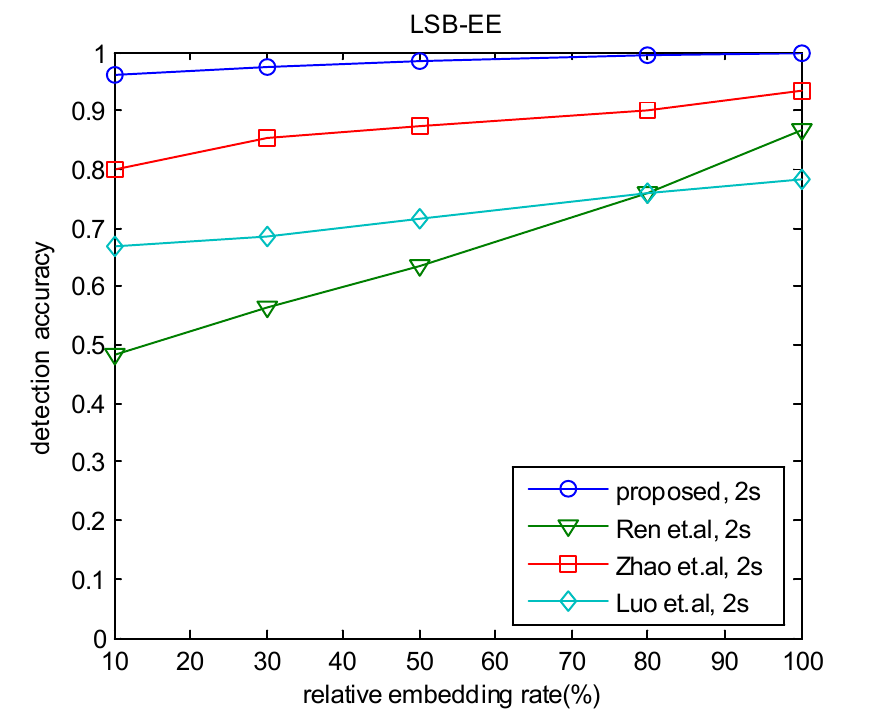}
\end{minipage}%
}%
\subfloat[]{
\begin{minipage}[t]{0.3\linewidth}
\centering
\includegraphics[width=2.2in]{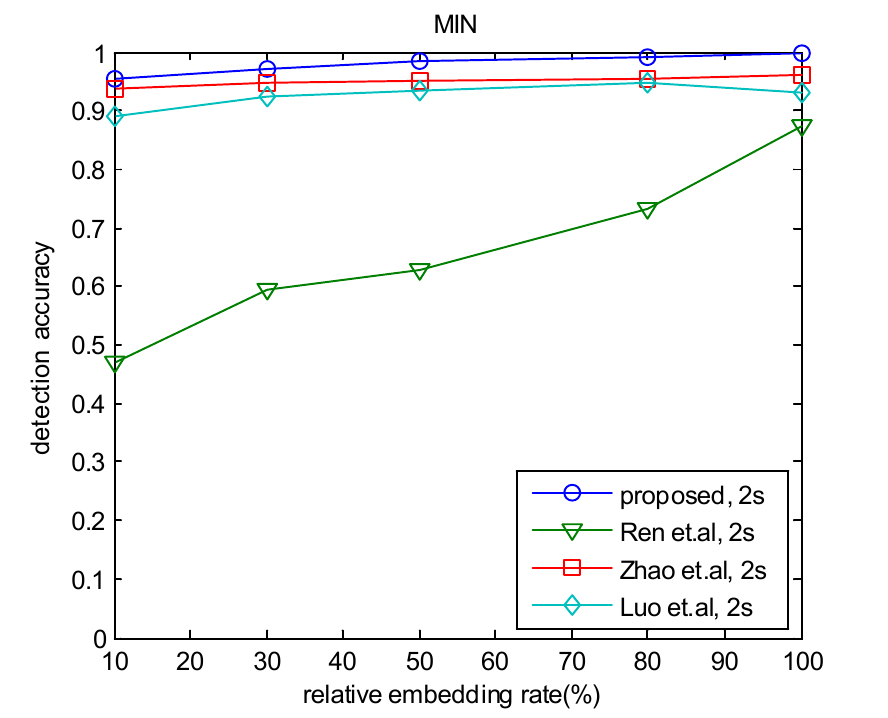}
\end{minipage}%
}%
\subfloat[]{
\begin{minipage}[t]{0.3\linewidth}
\centering
\includegraphics[width=2.2in]{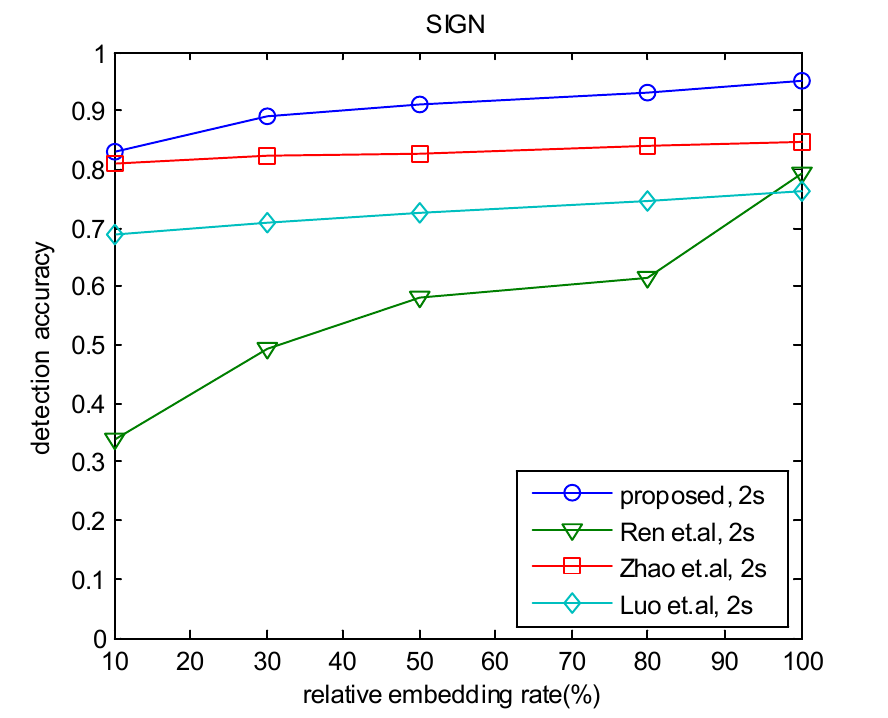}
\end{minipage}
}
  \caption{The detection accuracy of the  Spec-ResNet, Ren\cite{ren2017steganalysis}, Zhao\cite{wang2018cnn}, and Luo\cite{chen2017audio} steganalysis schemes for detecting existing AAC steganography schemes: (a) LSB-EE\cite{wang2010steganography}, (b) MIN\cite{wang2011steganography}, and (c) SIGN\cite{zhu2010sign}.}\label{fig_12}
\end{figure*}

The results of experiment \uppercase\expandafter{\romannumeral2} are shown in Table \ref{tab_2}. \textbf{Method} represents the steganalysis scheme, and \textbf{EA}, \textbf{AVERAGE} and \textbf{ACC} are as in Table \ref{tab_1}. Table \ref{tab_2} shows that the detection performance of the  Spec-ResNet steganalysis scheme is better than that of the three trained models in Table \ref{tab_1}, which shows that the feature combination of spectrograms with different window sizes can improve the performance of Spec-ResNet. Compared with other steganalysis schemes, the average detection accuracy of Spec-ResNet is higher than that of the compared schemes, including traditional hand-crafted schemes\cite{ren2017steganalysis} and CNN-based schemes\cite{chen2017audio,wang2018cnn}. In \cite{ren2017steganalysis}, the length of the audio samples is 20 seconds; however, Spec-ResNet still outperformed that scheme when applied to 2 s audio samples. Fig. \ref{fig_12} is a graphical representation of Table \ref{tab_2}.
 \begin{table*}[htbp]
   \centering
   \caption{The performance of various steganalysis schemes, Spec-ResNet, Ren\cite{ren2017steganalysis}, Zhao\cite{wang2018cnn}, and Luo\cite{chen2017audio}, in detecting existing AAC steganography schemes.}
   \begin{spacing}{1.35}
   \setlength{\tabcolsep}{3mm}{
     \begin{tabular}{ccccccccc}
     \toprule
     \multirow{1}[6]{*}{\textbf{Steganalysis Scheme}} &
     \multirow{1}[6]{*}{\textbf{EA}} & \textbf{TNR} & \multicolumn{5}{c}{\textbf{TPR}} & \multirow{1}[6]{*}{\textbf{ACC}}\\
     &   & \textbf{Cover} & \textbf{10\%} & \textbf{20\%} & \textbf{30\%} & \textbf{50\%} & \textbf{100\%} \\
     \midrule
     \multirow{2}[6]{*}{Spec-ResNet:Mixture} & LSB-EE\cite{wang2010steganography} & \multirow{2}[6]{*}{0.9428} & 0.9617 & 0.9758 & 0.9858 & 0.9938 & 1 & 0.9631 \\
      & MIN\cite{wang2011steganography} &   & 0.9554 & 0.9717 & 0.9858 & 0.9917 & 1 & 0.9619 \\
      & SIGN\cite{zhu2010sign} &   & 0.8283 & 0.8904 & 0.9108 & 0.9317 & 0.95 & 0.9245 \\ \hline
     \multicolumn{2}{r}{\textbf{AVERAGE \quad}} & \textbf{0.9428} & \textbf{0.9151} & \textbf{0.946} & \textbf{0.9608} & \textbf{0.9724} & \textbf{0.9833} & \textbf{0.9498} \\ \hline
     \multirow{2}[6]{*}{Ren\cite{ren2017steganalysis}} & LSB-EE\cite{wang2010steganography} & \multirow{2}[6]{*}{0.633} & 0.4825 & 0.5632 & 0.6345 & 0.7597 & 0.8661 & 0.6471 \\
      & MIN\cite{wang2011steganography} &   & 0.4694 & 0.5944 & 0.6291 & 0.731 & 0.8726 & 0.6462 \\
      & SIGN\cite{zhu2010sign} &   & 0.3384 & 0.4937 & 0.5814 & 0.6148 & 0.7933 & 0.5987 \\ \hline
     \multicolumn{2}{r}{\textbf{AVERAGE \quad}} & \textbf{0.633} & \textbf{0.4301} & \textbf{0.5504} & \textbf{0.615} & \textbf{0.7018} & \textbf{0.844} & \textbf{0.6307} \\ \hline
     \multirow{2}[6]{*}{Zhao\cite{wang2018cnn}} & LSB-EE\cite{wang2010steganography} & \multirow{2}[6]{*}{0.8819} & 0.7994 & 0.8525 & 0.8723 & 0.901 & 0.9342 & 0.8769\\
        & MIN\cite{wang2011steganography} &   & 0.9375 & 0.9479 & 0.9507 & 0.9531 & 0.9609 & 0.9160 \\
       & SIGN\cite{zhu2010sign} &   & 0.8098 & 0.8229 & 0.8281 & 0.8411 & 0.8463 & 0.8558 \\ \hline
     \multicolumn{2}{r}{\textbf{AVERAGE \quad}} & \textbf{0.8819} & \textbf{0.8489} & \textbf{0.8741} & \textbf{0.8837} & \textbf{0.8984} & \textbf{0.9138} & \textbf{0.8829} \\ \hline
     \multirow{2}[6]{*}{Luo\cite{chen2017audio}} & LSB-EE\cite{wang2010steganography} & \multirow{2}[6]{*}{0.8825} & 0.668 & 0.684 & 0.7161 & 0.7578 & 0.7839 & 0.8022 \\
       & MIN\cite{wang2011steganography} &   & 0.8919 & 0.9258 & 0.9336 & 0.9466 & 0.931 & 0.9041 \\
        & SIGN\cite{zhu2010sign} &   & 0.5039 & 0.5143 & 0.526 & 0.5378 & 0.5729 & 0.7067 \\ \hline
     \multicolumn{2}{r}{\textbf{AVERAGE \quad}} & \textbf{0.8825} & \textbf{0.6879} & \textbf{0.708} & \textbf{0.7252} & \textbf{0.7474} & \textbf{0.7626} & \textbf{0.8043} \\
     \bottomrule
     \end{tabular}}
      \end{spacing}%
   \label{tab_2}%
 \end{table*}%

\subsubsection{Experiment \uppercase\expandafter{\romannumeral3}}
To evaluate the detection performance of  Spec-ResNet for MP3s, MP3Stego is chosen as the detection object. The training procedure is the same as in \emph{Experiment \uppercase\expandafter{\romannumeral1}}. The audio duration is 5 s, and the three different window sizes are $N=512,\ 1024,\ or\ 2048$ (for $N=512,\ 1024,\ or\ 2048$, the dimensions of the feature matrix are $256\times 561, 512\times 424,$ and  $1024\times 214$, respectively). The samples in the \emph{MP3 audio data set} are 44.1 kHz samples. The training set consists of the spectrograms of 1500 audio samples in \emph{CDB\_MP3} and of the 1500 audio samples in \emph{SDB\_MP3}, and the test set is the same as the training set in the architecture. There is no intersection between the training set and the test set. The performance of Spec-ResNet is compared with other steganalysis schemes. Wang's method\cite{jin2017steganalysis} proposed a Markov feature to capture the correlations between the quantized MDCT coefficients (referred to as QMDCTs), which are destroyed by MP3stego's embedding behaviour. Luo's scheme \cite{chen2017audio} is compared for detecting MP3stego with an audio duration of 5 s, in which 8000 audio samples are chosen to train the CNN model and 2000 audio samples are used to test the trained model.

The detection performance of the  proposed  Spec-ResNet scheme for MP3Stego was compared with the other schemes proposed in \cite{jin2017steganalysis,chen2017audio}. The results are presented in Table \ref{tab_3}. \textbf{Mixture} represents the mixture classification model, which is trained on  three spectrograms with different window sizes, and \textbf{S\_len} is the sample length. Spec-ResNet outperforms the trained model based on a spectrogram with only one window size. The detection accuracy (ACC) of the proposed scheme, which is in bold, is higher than the compared steganalysis schemes based on traditional hand-crafted features\cite{jin2017steganalysis} or deep neural networks\cite{chen2017audio}. This means that the proposed  Spec-ResNet steganalysis scheme can also be applied to other audio coding standards, such as MP3, and can achieve high detection accuracy.

 \begin{table}[htbp]
   \centering
   \caption{The performance of the steganalysis schemes to detect MP3Stego\cite{petitcolas1998mp3stego}.}
   \begin{spacing}{1.35}
   \setlength{\tabcolsep}{0.5mm}{
     \begin{tabular}{cccccc}
     \toprule
        \textbf{Steganalysis Scheme} & \textbf{S\_len} & \textbf{EBR} & \textbf{TNR} & \textbf{TPR} & \textbf{ACC} \\
     \midrule
     Spec-ResNet:$N$=512 & 5 s & 6\% & 0.9960 & 0.9825 & 0.9892 \\
     Spec-ResNet:$N$=1024 & 5 s & 6\% & 1 & 0.9980 & 0.9990 \\
     Spec-ResNet:$N$=2048 & 5 s & 6\% & 0.9913 & 0.9980 & 0.9947 \\
     Spec-ResNet:Mixture & 5 s & 6\% & \textbf{1} & \textbf{0.9987} & \textbf{0.9993} \\
     Luo\cite{chen2017audio}& 5 s & 6\% & 0.9813 & 0.9759 & 0.9786 \\
     Wang\cite{jin2017steganalysis}& 10s & 10\% & $-$ & $-$ & 0.9250 \\
     \bottomrule
     \end{tabular}}
     \end{spacing}%
   \label{tab_3}%
 \end{table}%

\section{Conclusion}
In this paper, a general steganalysis scheme, Spec-ResNet, based on  spectrograms and deep residual networks is proposed. Spec-ResNet can be used to detect the steganographic schemes of the different embedding domains of AAC and MP3. Three conclusions can be drawn from this work: 1). A spectrogram represents the spectral relationships of time series of audio signals and can be effectively used as an analysis signal for audio steganalysis schemes to obtain more universal features. The spectrogram can be extended to  other audio analysis areas, such as audio forensics, to improve the classification performance. 2) Deep residual networks, which solve the problem of gradient disappearance, are suitable for extracting steganalysis features based on weak signal changes. 3) The fusion of training features from different scales improves the classification accuracy.

The experimental results show that the proposed scheme achieves good detection accuracy and generalizability. The scheme achieves a better detection accuracy for the three main AAC embedding domains, MDCT, scale factor and Huffman coding, and MP3Stego compared to other steganalysis schemes. To the best of our knowledge, this paper is the first time an audio steganalysis scheme based on spectrograms and deep residual networks has been proposed. The method in this paper can be applied to other types of audio classification works.

\section*{Acknowledgement}
The authors would like to thank Dr. Weiqi Luo for sharing the source code to their audio steganalysis scheme in \cite{chen2017audio}.
%
%

\ifCLASSOPTIONcaptionsoff
  \newpage
\fi



\bibliographystyle{IEEEtran}
\bibliography{reference}
\end{document}